# Tunable ultra-narrowband grating filters in thin-film lithium niobate

Alessandro Prencipe, Mohammad Amin Baghban† and Katia Gallo*

Department of Applied Physics, KTH Royal Institute of Technology, Roslagstullsbacken 21, Stockholm SE-106 91, Sweden

**ABSTRACT:** Several applications in modern photonics require compact on-chip optical filters with a tailored spectral response. However, achieving sub-nanometric bandwidths and high extinction ratios is particularly challenging, especially in low-footprint device formats. Phase shifted Bragg gratings implemented by sidewall modulation of photonic nanowire waveguides are a good solution for on-chip narrowband operation with reasonable requirements in fabrication and scalability. In this work we report on their implementation and optimization in thin film lithium niobate, a photonic platform that affords reconfigurability by exploiting electrooptic effects. The phase-shifted Bragg grating filters have a footprint smaller than 1 µm× 1mm and operate at telecom wavelengths, featuring extinction ratios up to 25 dB. We demonstrate transmission bandwidths as narrow as 14.4 pm (Q = 1.1 x $10^5$) and 8.8 pm (Q = 1.76 x $10^5$) in critically coupled structures and multi-wavelength Fabry-Perot configurations, respectively, in full agreement with theoretical predictions. Moreover, by taking advantage of the strong electrooptic effect in lithium niobate, in combination with the tight light confinement of nanophotonic wires and the ultranarrow spectral resonances of optimized grating structures, we demonstrate a tunability of 25.1pm/V and a record modulation of the filter transmission amounting to 1.72 dB/V at CMOS voltages. The results pave the way for reconfigurable narrowband photonic filters in lithium niobate with small footprint and low consumption, to be exploited towards on-chip quantum and nonlinear optics, as well as optical sensing and microwave photonics.

KEYWORDS:  Bragg grating, microcavity, nanophotonics, lithium niobate, electro-optics.

Bragg gratings (BG) are essential components for a wide variety of devices and applications (1, 2), encompassing optical filters and lasers (3, 4), add-drop multiplexing, temporal imaging and dispersion engineering in telecom systems (5-7), integrated optical sensors (8, 9) and advanced photonic and microwave signal processing devices (10, 11). On-chip integrated Bragg gratings are typically realized by implementing a periodic sidewall corrugation of nanophotonic waveguides (12, 13). Such width-modulation can be introduced directly in the waveguide lithographic fabrication step, allowing easy and repeatable realization of ultrasmall footprint devices. However, achieving sub-nanometric bandwidths with uniform Bragg gratings (BG) remains challenging, as it requires extremely shallow sidewall corrugations, motivating the quest for alternative approaches (12, 14, 15). One of them was envisaged already in the very early days of integrated optics and involves the use of phase shifted Bragg gratings (PSBG), implemented via engineered defects in BG structures (16). Optimized PSBGs yield ultranarrow bandwidths and high contrasts in low footprint devices (17, 18).

The possibility of making such photonic filters reconfigurable is also very attractive. Tunability in BG devices can be introduced by means of thermal, strain or carrier injection effects (11, 19, 20). Nevertheless, these approaches may suffer from high power consumption and enhanced optical losses. Such drawbacks can be eliminated by resorting to electro-optic materials in nanophotonic formats, a prospect made viable by the recently emerging thin-film lithium niobate on insulator (LNOI) platform (21). LNOI photonic integrated circuits combine the advantages of nanophotonics with the unique properties of LN, enabling advanced nonlinear and electro-optic on-chip functionalities (22, 23). High quality LNOI nanocavities and modulators have been demonstrated with photonic crystal nanobeam waveguides (24, 25). An alternative approach, based on sidewall-modulated photonic wires similar to silicon-on-insulator (SOI), has been simultaneously pursued for integrated Bragg grating filters and modulators in LNOI (26-29). The sidewall modulated grating technology affords enhanced mechanical stability as well as good reproducibility, scalability and CMOS compatibility of the device fabrication process. It has yielded BG filters featuring rejection bandwidths of ~10 nm at telecom wavelengths, with high extinction ratios and an electro-optic (voltage-to-wavelength) tunability coefficient $\sigma_\lambda = \partial\lambda/\partial V = 23.4$ pm/V at the edge of the bandgap. Moreover, recent preliminary results obtained by our group and others, indicate the potential of PSBG LNOI structures for integrated filters with sub-nanometric bandwidths and electro-optic tunability (30-33).

In this work we present a comprehensive experimental study, correlated by theoretical analyses, on sidewall-modulated PSBG devices in LNOI nanophotonic wires as shown in Figure 1a, including both π-phase-shifted (quarter-wavelength) and long-cavity (Fabry-Perot) configurations, enabling single and multi-wavelength operation in the telecom range. We demonstrate bandwidths as narrow as 8.8 pm, with minimal transmission penalty, in full agreement with the theoretical expectations. Furthermore, by taking advantage of the fine spectral features of the PSBG and of the enhanced electro- optic effects using the $d_{33}$ coefficient in LNOI nanophotonic wires with integrated side-electrodes ($\sigma_\lambda = 25.1$ pm/V), we achieve a record tunability of the device optical transmission, amounting to $\sigma_T = \partial T / \partial V = 1.725\ dB/V$. This corresponds to a measured overall transmission on/off ratio as high as 6.9 dB for an applied voltage of 4 V. Besides providing further confirmation for the degree of maturity reached by the LNOI BG technology platform, the results pave the way for its effective deployment for coherent spectral manipulation of photons in ultrasmall-footprint and low-consumption devices for a broad range of applications, spanning from reconfigurable optical signal processing in telecommunications to programmable quantum optics, microwave photonics and optical sensing.

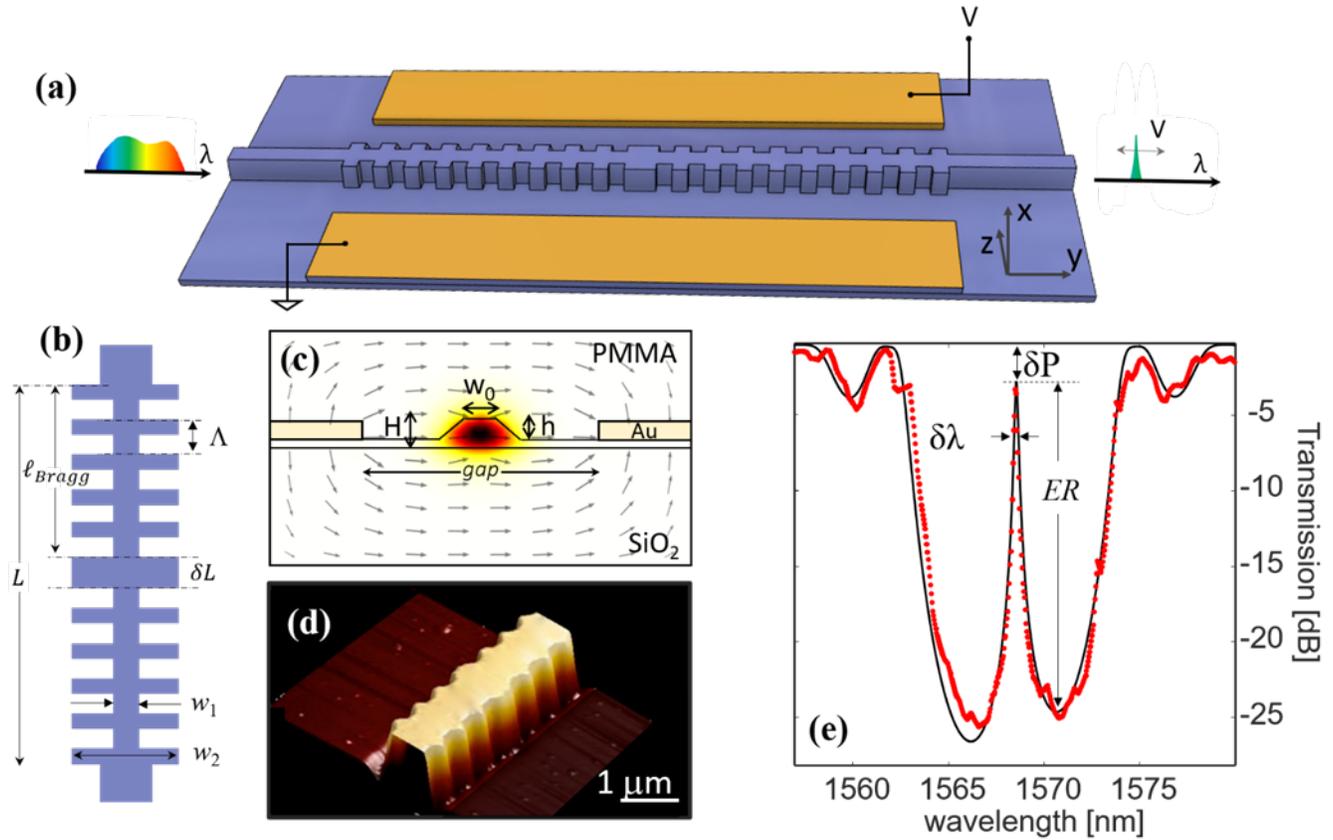

Figure 1. (a) Schematic of a phase-shifted Bragg grating (PSBG) device implemented with sidewall-modulated waveguide technology in LNOI. x, y and z are the LN crystal axes. (b) Top view highlighting key PBSG waveguide parameters. (c) Cross-sectional view of the waveguide with computed $TE_{00}$-mode electric field distribution at $\lambda = 1550\ nm$, for $w_0= 650\ nm$, $H = 500\ nm$, $h = 300\ nm$ and electrostatic field when the gap between the electrodes is 4 μm. (d) Detail of an etched PSBG structure imaged by atomic force microscopy. (e) Transmission spectrum of a π-PSBG LNOI device with $w_0$ = 450 nm, $\delta w$ = 250 nm, $H$ = 500 nm, $h$ = 360 nm, period $\Lambda = 435\ nm$, $DC$ = 0.78 and propagation losses $\alpha_0 = 2.9 dB/cm$, yielding $\lambda_0 = 1568\ nm$, $\delta\lambda$ = 142 pm, $\delta P = 2.43 dB$, $ER = 22.2 dB$. Red dots: experimental data; Black solid line: simulations.

## DEVICE OVERVIEW

The PSBG device structure is sketched in Figure 1a. It is implemented with nanophotonic waveguides fabricated in X-cut LNOI wafers (NANOLN Ltd) and designed for operation with quasi-$TE_{00}$ modes at telecom wavelengths. Integrated BGs are realized by sidewall modulating the nanowire top width between the values $w_1 = w_0-\delta w$ and $w_2 = w_0+\delta w$, with period $\Lambda$, duty cycle $DC$ and total device length $L$ along the propagation direction ($Y$). The device fabrication process involves electron beam lithography followed by dry etching and is described in the Methods section. As shown in Figure 1b, a localized defect, consisting in an aperiodic element of length $dL$, is introduced between two BG sections of equal length: $\ell_{Bragg}=(L-dL)/2$, to realize a PSBG device.

Figure 1c highlights key geometrical parameters of the waveguide, namely the top width ($w_o$), the height ($h$) of the photonic wire and the original LNOI thickness ($H$), with its top and bottom cladding layers (PMMA and SiO$_2$, respectively). The figure also highlights the gap size between the gold electrodes used to tune the spectral response. The transverse field distribution of the $TE_{00}$ optical guided mode at a wavelength of 1550 nm, computed with a commercial eigenmode solver (COMSOL), together with the electrostatic field distribution (represented by the arrows) obtained by applying a unitary voltage across the electrodes.

For a given BG waveguide geometry, yielding an average $TE_{00}$-mode index $\bar{N}$ at a given wavelength $\lambda$, the sidewall modulation results in the opening of a photonic bandgap around the Bragg wavelength $\lambda_0 = 2\bar{N}_0\Lambda$. The PSBG defect $dL$ introduces an additional phase shift $\varphi(\lambda) = 4\pi\ \bar{N}(\lambda)\ (dL - \Lambda/2)/\lambda$ between the forward and backward propagating waves at a generic wavelength $\lambda$. For $dL=\Lambda$ a sharp transmission peak appears at the center of the photonic bandgap. This structure, corresponding to $\varphi = \pi$ at the Bragg wavelength, represents a π-phase-shifted Bragg grating (π-PSBG). A detail of its implementation in LNOI is shown in Figure 1d. Figure 1e illustrates the typical transmission spectrum of a π-PSBG and compares theory (black solid line) with measurements (red markers), highlighting key figures of merit for the device performance as a narrowband filter, namely the 3-dB bandwidth $\delta\lambda$, extinction ratio $ER$ and power penalty $\delta P$ of the transmission peak at the Bragg wavelength $\lambda_0$. Transmission peaks can be introduced in the bandgap also by longer defects, $dL \gg \Lambda$, which sustain higher order resonances, i.e. $\varphi = \pi(1+2m)$ with integer $m$. For sufficiently long $dL$, the bandgap accommodates multiple peaks, whose spectral separation scales as $\sim 1/dL$, similarly to Fabry-Perot etalons. Such structures are referred to in what follows as long-cavity devices.

## QUALITY FACTORS

In this section we introduce a simple 1D-cavity model providing intuitive insights and guidelines for device optimization. The PSBG structure (Fig. 1b) is considered as a lumped cavity of length $L_{cavity}$ centered around the defect $dL$ and comprised between two mirrors. The distributed nature of the latter implies a non-zero penetration depth of the mode field into the sections $\ell_{Bragg}$. The latter effect is particularly relevant for π-PSBG structures but can be neglected for long-cavity PSBG devices, to which the further approximation $L_{cavity} \sim dL$ can be applied.

Notwithstanding the value of $dL$, the $Q$-factor of the cavity stems from the contribution of two terms: one arising from losses ($Q_\alpha$) and the other from coupling into the cavity ($Q_\kappa$). The former is associated with the loss coefficient $\alpha$ and independent of the device length, while the latter is determined by the reflectivity of the Bragg mirrors, which exhibits an exponential dependence on the grating coupling coefficient $\kappa$ and on the grating length $\ell_{Bragg}$ (3, 34).

The transmission bandwidth $\delta\lambda$ of the peak at $\lambda_0$ is directly related to the loaded Q-factor of the PSBG cavity and can be expressed in terms of $Q_\alpha$ and $Q_\kappa$ as (16, 35):

$$\frac{\delta\lambda}{\lambda_0} = \frac{1}{Q} = \frac{1}{Q_\kappa} + \frac{1}{Q_\alpha} \quad (1)$$

Accordingly, the spectral features of the cavity mode are determined by an interplay between waveguide losses and grating strength (36). For a given $\alpha$, the intrinsic Q factor is fixed and equals $Q_\alpha = \pi/(\alpha\Lambda)$ at the Bragg wavelength. The bandwidth of the transmission peak at $\lambda_0$ is then minimized by maximizing $Q_\kappa$.

For a π-PSBG, increasing the coupling strength, to maximize $Q_\kappa$, increases the Bragg mirror reflectivity. However, this makes the injection of optical power into the cavity less efficient, which translates into an increasing power penalty $\delta P$(37). An optimum is attained at $Q_\kappa = Q_\alpha$, a condition known as *critical coupling*, corresponding to the low-power penalty and smallest achievable bandwidth limit $\delta\lambda_c = 2\alpha\Lambda\lambda_0/\pi$. Critical coupling implies striking a fine balance between coupling ($\kappa$) and propagation losses ($\alpha$) by careful adjustment of the fabrication conditions, as discussed in the next section.

An alternative approach to minimize the bandwidth involves using long-cavity ($dL \gg \Lambda$) PSBG devices where the condition $Q_\kappa \gg Q_\alpha$ is afforded by increasing $L_{cavity} \sim dL$, without affecting the mirror reflectivity. With a lumped Fabry-Perot cavity model, the transmission linewidth is:

$$\delta\lambda = \frac{\lambda_0^2}{2\pi\bar{N}}\left(\alpha - \frac{\ln R}{L_{cavity}}\right) \quad (2)$$

(34), where $R = \tanh^2(\kappa\ell_{Bragg})$ is the Bragg mirror reflectivity. Thanks to the presence of $L_{cavity}$ at the denominator, Q-factors approaching the intrinsic (unloaded cavity) limit can be attained by increasing the length of the phase-shifting segment $dL$, without the need for a critical adjustment of the grating strength to match intrinsic losses. This approach increases the device footprint and simultaneously reduces the cavity free spectral range (FSR), $\Delta\lambda_{FSR} = \lambda_0^2/(2L_{cavity}N_g)$, resulting in the appearance of several peaks in the transmission spectra for increasing $L_{cavity}$.

## TRADEOFFS and DEVICE OPTIMIZATION

A detailed discussion of the tradeoffs and optimization of the π-PSBG devices is provided in SI. Here we summarize its main findings with concern to the impact of two key parameters: the sidewall modulation amplitude $\delta w$ and the overall device length $L$. The evolution with $\delta w$ of the figures of merit of the PSBG response is shown in Figure 2.

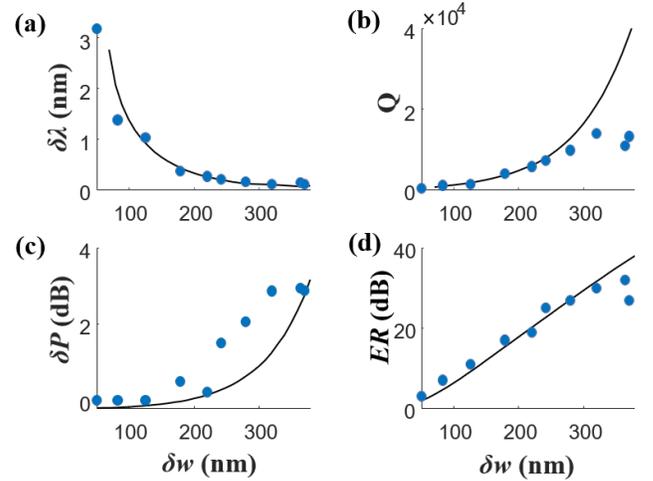

Figure 2. Figures of merit for the π-PSBG filter response plotted as a function of the grating modulation depth $\delta w$: (a) 3dB transmission bandwidth, (b) cavity Q-factor, (c) peak power penalty and (d) extinction ratio, for a device set with $w_0 = 450\ nm$, $L = 217\ \mu m$, $\Lambda = 435\ nm$, $H = 500\ nm$, $h = 360\ nm$. Filled circles: measurements. Solid lines: simulation results for $\alpha_0$ = 2.9dB/cm (loss value measured on unmodulated waveguides).

Considering otherwise identical PSBG devices, Fig. 2a and 2b plot the peak transmission bandwidth and loaded quality factor, respectively, for $\delta w$ varying from 50 to 370 nm, while Fig. 2c and 2d illustrate the evolution of the power penalty and extinction ratio, respectively. The circles are experimental data, while the solid lines are numerical predictions obtained with the model described in Methods, under the assumption of a constant loss coefficient, equal to the one of unmodulated waveguides, i.e. $\alpha_0 = 2.9\ dB/cm$ for $w_0 = 450\ nm$, in this case. For values of $\delta w$ up to 240 nm, the experiments feature a monotonic bandwidth decrease (Q factor increase) which is fully consistent with theory.

However, beyond that point the transmission bandwidth appears to saturate at $\delta\lambda \sim 115$ pm (Fig. 2a), corresponding to $Q = 1.37 \times 10^4$ (Fig. 2b). This is in contrast with theory (solid line), which predicts a minimum bandwidth of 26 pm at critical coupling when $\alpha_0 = 2.9\ dB/cm$. The experimental results also indicate significant power penalties $\delta P$ (indication of non-critical coupling) and a degradation of the peak extinction ratio $ER$. The performance degradation observed at high grating modulation depths ($\delta w > 240\ nm$) exhibits similarities with effects previously reported for other nanophotonic platforms, such as SOI, and is compatible with the insurgence of additional non-negligible scattering losses in the PSBG devices for $\delta w > 200$ nm. For a quantitative evaluation of the additional loss term $\alpha_s$ induced by the grating, we used our simulation tools to perform numerical fits on the experimental spectra with $\alpha_s$ as an adjustable parameter. We could then recover an excellent agreement between theory and experiments even at larger values of $\delta w$, yielding inferred scattering losses ranging between 3 and 10dB/cm, for $\delta w$ between 240 and 370nm (Suppl. Inform.).

The detrimental impact of scattering losses sets an upper bound to the possibility to attain the narrowest linewidths uniquely by increasing $\delta w$. However, the actual dependence of $Q_\kappa$ on the normalized quantity $\kappa \ell_{\text{Bragg}}$, affords an additional route for device optimization, relying on an adjustment of $\ell_{\text{Bragg}}$ and hence the device length $L$. This approach is illustrated in Figure 3.

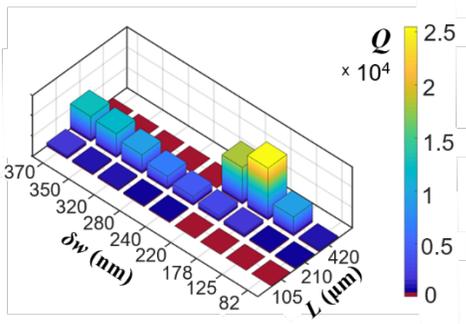

Figure 3. Experimental 2D tomography of the loaded quality factors of π-PBSG waveguides as a function of sidewall corrugation depth ($\delta w$) and overall device length ($L$), performed with $w_0 = 650\ nm, H = 500\ nm, h = 360\ nm$ and $\Lambda = 420\ nm$.

The 2D histogram of Figure 3 shows the loaded $Q$ factors measured for a set of 27 waveguides, made on the same chip and encompassing three different values of $L$ (105 - 420 nm) and nine of $\delta w$ (82 – 370 nm). The shortest gratings appear to yield cavities which are strongly undercoupled for all modulation depths. Doubling the device length ($L = 210\ \mu m$) yields a monotonic increase of the $Q$ factor with $\delta w$. However, it does not reach critical coupling. Ultimately, with a further increase of the device length to $L = 420\ \mu m$, critical coupling regime is achieved for $\delta w < 200$ nm, which minimizes also the scattering loss penalty. Critical coupling for $L = 420\ \mu m$ is apparent from the non-monotonic trend of $Q$ as a function of $\delta w$. The highest bar in the 2D map (Q = $2.6 \times 10^4$) is achieved for $\delta w = 178\ nm$ and corresponds to a measured transmission bandwidth $\delta\lambda = 59\ pm$, with a power penalty of 3 dB and an extinction ratio of 25 dB.

Besides a careful choice of $\delta w$ and $L$, the final optimization of both π-PSBG and long-cavity devices involved the overall minimization of waveguide propagation losses on the LNOI platform. This was achieved by improving the waveguide design and fabrication, while targeting a nanowire width $w_0 \sim 650\ nm$ and an etching depth $h \sim 300\ nm$, to achieve a good compromise between modal confinement, propagation losses, grating coupling strengths and device footprint. Ultimately, we settled for $w_0 = 640\ nm$ and $h \sim 310\ nm$, yielding $\alpha = 1.5$ dB/cm. This corresponds to a theoretical limit for the loaded Q factor in critically π-PSBG devices amounting to $Q_\pi = 1.1 \times 10^5$. Long cavity devices can further improve on this upper theoretical limit by a factor of two.

## NARROWBAND TRANSMISSION

The best results in terms of narrowband devices are highlighted in Figure 4, with reference to π-PSBG (Figure 4a) and long cavity PSBG (Figure 4b) devices.

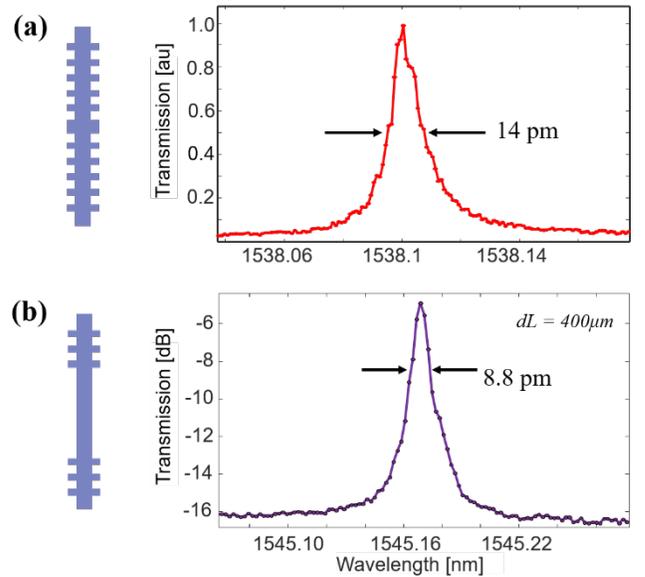

Figure 4. (a) π-PSBG device schematics and measured transmission peak for a PSBG with $H = 500nm, h = 310\ nm, w_0 = 640\ nm, \delta w = 100\ nm, \ell_{Bragg} = 331\mu m$ and $dL = \Lambda = 425nm$. (b) Long cavity PSBG device schematics and measured transmission peak for a PSBG with H = 500nm, $h = 310\ nm$, w0 = $640\ nm, \delta w = 100\ nm, \Lambda = 425nm, \ell_{Bragg} = 273\ \mu m$ and $dL = 400\mu m$.

Critically coupled devices with loaded Q factors in excess of $10^5$ were consistently achieved for the former, in agreement with the theoretical limit for π-PSBG devices. An example is provided in Figure 4a, showing the spectrum of the transmission peak measured in a 680μm-long π-PSBG with a modulation depth $\delta w$ = 100 nm, exhibiting a bandwidth of 14.4 pm (Q = 1.06 ×$10^5$), with an ER of ~ 19dB. Long-cavity PSBG designs yielded even narrower bandwidths, with experimental values in full agreement with the predictions of Eq. (1). This is illustrated by Figure 4b showing the central Bragg resonance peak of a long-cavity PSBG device with $dL = 400\ \mu m$, featuring a bandwidth as narrow as $\delta\lambda = 8.8\ pm$ (Q ~ $1.76 \times 10^5$) and an ER of 20dB. To the best of our knowledge, this is the narrowest peak ever reported on this kind of 1D resonators on LN.

The full extent of the multi-peaked spectral response measured in the long cavity PSBG devices is shown in Figure 5, where we show the evolution of the PSBG transmission spectra for four different cavity lengths $dL$, comprised between 100 μm and 400 μm. As $dL$ is increased, the number of transmission peaks within the bandgap increases as a result of a progressive decrease in the FSR. For the shortest cavity, i.e. $dL$ = 100 μm, the FSR is almost identical to half the width of the photonic bandgap ($FSR = 2.25\ nm$), yielding a single peak around the Bragg wavelength, with a transmission bandwidth $\delta\lambda = 10.4\ pm$ ($Q \sim 1.49 \cdot 10^5$). On the other end, for $dL$ = 400 μm, the FSR reaches a value of 1.06 nm, yielding three transmission peaks located well within the photonic bandgap. Each of them features comparable bandwidths to the one highlighted in Figure 4b. The smallest power penalty is obtained for $dL$ = 100 μm and amounts to $\delta P = 1.95\ dB$.

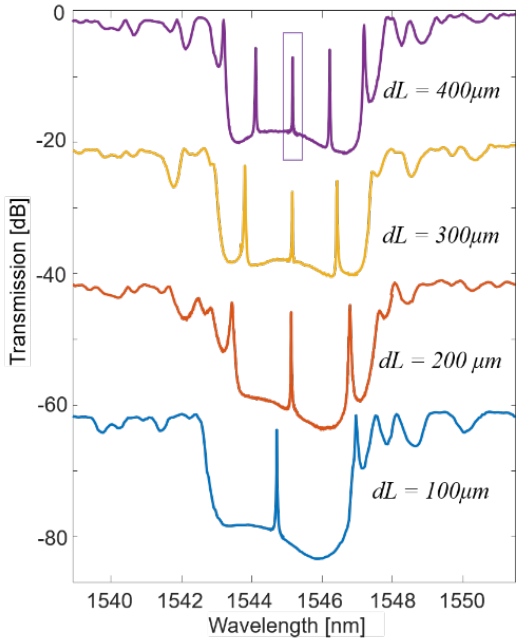

Figure 5. Full transmission spectra measured on the device of Fig. 4d (top spectrum) and for different values of $dL$ (100, 200, 300μm), all other device parameters being the same. The peak in the box is the one shown in Figure 4d.

## ELECTRO-OPTIC TUNING

The transmission peak can be tuned through the additional application of a voltage to electrodes deposited by the sides of the LNOI waveguide (27, 38). For our typical PBSG waveguide geometry ($w_0 = 640\ nm$, $H = 500\ nm$ and $h = 310\ nm$) the optimal gap between the electrodes is found to be 4 μm (see also Suppl. Information). The computed electrostatic field lines and the transverse profile of the fundamental guided optical mode are shown in Figure 1c. When a positive voltage is applied to the +Z side of the LN rib, the Bragg resonance wavelength experiences a red shift, as detailed in Suppl. Information. A record tunability of $\sigma_\lambda = \partial\lambda/\partial V = 25.1\ pm/V$ was measured on optimized π-PSBG devices, with 670μm-long side electrodes. Long cavity devices with $dL = 400\ \mu m$ yielded a value of 14.3 $pm/V$, comparable to previous reports (38).

The optimal tuning conditions for π-PSBG devices were obtained with electrodes running along the full length $L$ of the structure, as sketched in Figure 6a. This induces a resonance shift by accumulating an electro-optic refractive index change in the distributed Bragg mirrors, rather than in the submicrometric cavity ($dL << \ell_{Bragg}$). The plots in Figure 6a illustrate the full spectral responses of a π-PSBG measured for voltages of $-15, 0$ and $+15\ V$ on the side electrodes (yellow, blue and red curve, respectively). The voltage simultaneously shifts the mid-gap transmission peak and the edges of the photonic bandgap, as in ordinary BG devices (11, 27). However, as apparent from the curves, the BG features of the spectrum far from the resonance peak get deformed as one applies a voltage to the electrodes, whereas the shape of the PSBG peak stays the same. Similar distortions have been observed elsewhere in BG devices operating at the band edges (38) and limit the achievable tuning performance especially at low operating voltages. Such spectral perturbations are essentially absent in the tuning of the transmission peak inside the photonic bandgap. Furthermore, the ultranarrow-bandwidth of PSBG devices makes them particularly advantageous for low-voltage electro-optic operation. A relevant figure of merit in this context is the transmission contrast coefficient $\sigma_T = \partial T/\partial V$, that is the change in device transmission per unit voltage applied to the tuning electrodes, which for the π-PSBG device of Figure 6a amounts to 0.6 dB/V.

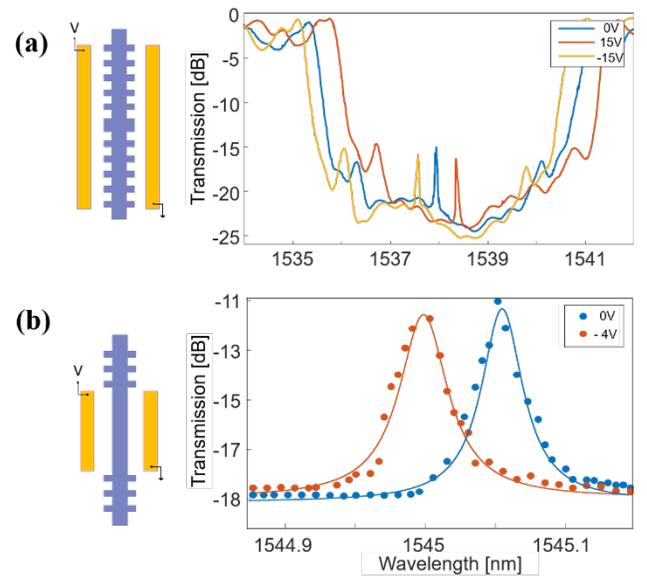

Figure 6. Electrooptic tunability. (a) electrode configuration and measured transmission spectra of a π-PSBG device with external voltages of 0 and ± 15V. (b) Electrode configuration and transmission peak tuning with a voltage of 4V in a long cavity PSBG device ($L_{cavity} = 400\mu m$). The transmission change amounts to 6.9 dB.

Optimized long cavity devices afford a further improvement to this figure of merit as shown in Figure 6b. In this case the EO tuning is applied only to the $L_{cavity}$ segment. This shifts the transmission peaks in the photonic bandgap without affecting the spectral location of the latter.

The high-resolution spectral plots in Figure 6b demonstrate the experimental tuning of the transmission peak at a voltage of **4V**. The PSBG peak shifts by 57 pm with negligible spectral distortion, and a transmission change of 6.9 dB is achieved at the original Bragg wavelength λ = 1545.06 nm. This corresponds to a spectral tunability $\sigma_\lambda = 14.3 \text{ pm/V}$ and a transmission modulation efficiency $\sigma_T = 1.725 \text{ dB/V}$. The latter represents an improvement of more than one order of magnitude with respect to previous results achieved in static tuning of LNOI Bragg gratings ($\sigma_T \sim 0.12 \text{ dB/V}$) (27).

## CONCLUSIONS

We reported a systematic study on phase shifted Bragg gratings (PSBG) in thin film lithium niobate considering both quarter-wavelength (π-PSBG) and long-cavity configurations. A full mapping of the waveguide and Bragg grating parameter space was performed with simulations and experiments in devices fabricated on LNOI chips through a scalable process involving electron beam lithography and dry etching. This allowed us to identify key elements for device optimization and achieve ultra-narrowband operation and critical coupling in π-PSBG devices. The study enabled also to evaluate the scattering losses associated to deep grating modulations. By choosing relatively shallow modulations ($\delta w \sim 100 \text{ nm}$) we achieved experimentally the theoretical limit for critically coupled π-PSBG, measuring transmission bandwidths of 14.4 $pm$ ($Q \sim 10^5$) on devices with a footprint of only $1 \times 680$ μm². Good agreement between theory and experiments was also demonstrated for multi-wavelength resonant devices implemented with long cavities ($\delta L \sim 100 - 400 \text{ μm}$), yielding bandwidths of 8.8 $pm$ and loaded Q-factors of $1.76 \times 10^5$. Finally, by exploiting the electro-optic effect, we achieved a tunability of the transmission wavelength of 25.1pm/V and an optical transmission contrast at the Bragg wavelength of ~7 dB for voltages fully compatible with CMOS electronics. The combination of ultranarrow bandwidths achievable with PSBGs and electro-optic tunability in LN paves the way for low power consumption CMOS-compatible devices for electro-optic switching and modulation in telecommunication systems as well as efficient and fast photon manipulation in integrated quantum photonics. Moreover, the excellent agreement attained between theory and experiments demonstrates the technological maturity of the PSBG LNOI platform and holds promise for the implementation of more advanced functionalities for spectral shaping and tuning and their use for dispersion compensation as well as nonlinearity enhancement and frequency combs with $\chi^{(2)}$ nonlinearities (19, 39). The small footprints and low-voltage operation achieved in these devices as well as the scalability of their fabrication process might also be advantageously exploited towards perspective developments of microwave photonics on LNOI as well as the implementation of programmable and reconfigurable nanophotonic devices for e.g. multispectral sensing, neuromorphic and quantum computing(11, 40-42).

## METHODS

The fabrication process of Ref. (26) was optimized for improved reproducibility and to reach fine patterning resolutions with deeper LNOI etching. All the integrated nanophotonic components were simultaneous defined on chip by a single-step electron beam exposure (Raith Voyager, acceleration voltage 50 kV), patterning a resist layer (ma-N2400) spun on a chromium layer deposited on commercial X-cut LNOI chips (NanoLN Ltd). The chromium layer was patterned by $Cl_2/O_2$ reactive ion etching (Oxford Plasmalab 100) and used as a hard mask for $Ar^+$-ion milling of the underlying LN film. The process yielded nanowire waveguides with sidewall angles of 55-60 degrees. The tuning electrodes were patterned by liftoff of a 50 nm-thick Au layer with a 10 nm-thin Cr adhesion layer. The final LNOI devices were clad with a 2 μm-thick layer of PMMA (MicroChem 950) baked at 170°C (30, 31).

The optical characterizations were performed by coupling light from single mode optical fibers at telecom wavelengths into the LNOI chip with integrated LNOI grating couplers and tapers enabling selective excitation of the fundamental $TE_{00}$ mode in the PSBG nanowires. As in previous work (26), we used a tunable continuous-wave laser (Yenista T100S) as a source for spectrally-resolved measurements recording the device throughput off-chip with a fiber-coupled power meter (Newport Model 2931-C) synchronized with the tunable laser.

Numerical analyses of the PSBG response in the wavelength range of interest ($\lambda$ = 1500 - 1600 nm) were performed by using a commercial vectorial eigenmode solver (COMSOL Wave Optics) for the waveguide modal analysis and a couple-mode theory approach to determine the overall device transmission. The simulations were correlated to the waveguide {$w_0$, $h$, $H$} and grating {$\delta w$, $\Lambda$, $dL$} fabrication conditions, as detailed in Suppl. Inform. For each device design, the BG phase-mismatch $\Delta\beta$ was computed as a function of wavelength $\lambda$ from the average effective index $\bar{N}$ of the $TE_{00}$ mode in the sidewall-modulated waveguide. Similarly, the coupling constant $\kappa$ was determined from the mode-grating transverse overlap, assuming a rectangular modulation between the values $w_0 \pm \delta w$, with small adjustments to the grating duty-cycle (in the range 70-82%) to match the experiments and account for the non-ideality and rounded edges of the fabricated grating profiles. The compound response of the Bragg grating and phase-shifting sections was simulated with a transfer matrix approach, using a coupled-mode formalism in guided-wave configuration (2, 34). The model related the complex envelopes of the forward ($A$) and backward ($B$) propagating waves at the input ($A_0$ and $B_0$) and output ($A_L$ and $B_L$) of the overall structure, through a matrix relationship:

$$\begin{bmatrix} A_L(\lambda) \\ B_L(\lambda) \end{bmatrix} = M_{PSBG}(\lambda) \begin{bmatrix} A_0(\lambda) \\ B_0(\lambda) \end{bmatrix} \quad (6)$$

where $M_{PSBG} = M_{BG}M_{PS}M_{BG}$ and the transfer matrices $M_{BG} = M_{BG}(\Delta\beta, \kappa)$ and $M_{PS} = M_{PS}(\varphi)$ are determined by the analytical solutions for the uniform Bragg grating of length $\ell_{\text{Bragg}}$ and the central phase shifting element of length $dL$. Further details are provided as Suppl. Information.

## ASSOCIATED CONTENT

The Supporting Information contains additional content on: modelling of the device optical response (par. 1), theory and experiments on electrostatic tuning (par. 2), PSBG design trade-offs (par. 3), scattering loss evaluation and detailed comparison of experimental and theoretical responses (par. 4) and estimates of the effective cavity length (par. 5).

## AUTHOR INFORMATION

### Corresponding Author
* Corresponding author: Katia Gallo. gallo@kth.se

### Present Addresses
† Currently with AFRY AB, Frösundaleden 2A, Solna SE-169 99, Sweden.

### Author Contributions
K. G., A. P. and M.A.B. conceived the study, A. P. fabricated and characterized the devices, with contributions by M.A.B. A.P. and K. G. performed the design, modelling and data analysis and wrote the manuscript. All authors approved the final version of the manuscript.

## ACKNOWLEDGMENTS

We acknowledge financial support from the Knut and Alice Wallenberg Foundation through grant nr. 2017.099 and from the Wallenberg Center for Quantum Technology (WACQT), as well as from the Swedish Research Council through grant no. 2018-04487 and the research environment Optical Quantum Sensing (OQS, grant nr. 2016-06122). The fabrication has been carried out in the Albanova NanoLab facilities in Stockholm and the valuable technical support of its staff, particularly Erik Holmgren and Adrian Iovan, is also gratefully acknowledged.

# Low Voltage tunable Phase shifted Bragg grating filters in thin film Lithium Niobate – SUPPLEMENTARY INFORMATION

Alessandro Prencipe, Mohammed A. Baghban and Katia Gallo*

Department of Applied Physics, KTH Royal Institute of Technology, Roslagstullsbacken 21, Stockholm SE-106 91, Sweden

*KEYWORDS nanophotonics, tunable nanocavity, tunable ultranarrow filters, Bragg grating, thin film Lithium Niobate.*

Content

## 1. NUMERICAL MODELING – FULL DESCRIPTION

This section describes the main theoretical concepts and modelling tools which were used to simulate the device response and correlate the latter to actual fabrication conditions.

Coupled mode theory.

The spectral response of each device is determined by the geometrical properties of the unperturbed waveguide $\{w_0, H, h\}$ and of the grating modulation $\{dw, L, DC\}$ and was computed by a numerical approach which combined a transfer matrix method with guided-wave coupled-mode theory (1, 2).

For each wavelength λ in the range of interest (1500 - 1600 nm) and a given design $\{w_0, H, h, dw\}$, a vectorial eigenmode solver (COMSOL Wave Optics) was used to perform a modal analysis of three relevant waveguide geometries, corresponding to the waveguide widths $w_0$, $w_1 = w_0 - \delta w$ and , $w_2 = w_0 + \delta w$. The Bragg grating phase-mismatch was then computed as (1, 2):

$$\Delta\beta(\lambda) = \frac{4\pi}{\lambda} N_{av}(\lambda) - i\alpha - \frac{2\pi}{\Lambda}$$

where $N_{av}$ is the grating-averaged $TE_{00}$-mode effective index and $\alpha$ the waveguide power-attenuation coefficient. The coupling constant $\kappa$ of the Bragg grating was computed using the transverse field profile $F_0(x,z)$ of the z-polarized component of the fundamental mode obtained from the modal analysis of the unmodulated waveguide (width $w_0$). To do this, we applied the coupled-mode formalism for guided-waves, with field distributions normalized for unitary power flow (1, 2). By adopting a perturbative approach, the periodic perturbation of the dielectric permittivity associated with the sidewall modulation of the Bragg grating was expressed as: $\Delta\varepsilon = 2n_0\delta n$, where $n_0 = n_0(x, y)$ is the refractive index profile of the unperturbed waveguide and $\delta n = \delta n(x, y, z) = \delta n(x, y + \Lambda, z)$ is the index perturbation of the sidewall grating. The latter is further expanded with a Fourier series development of which only the first term was retained (1st order Bragg resonance), yielding: $\delta n = \delta n_1(x, z) \cdot \exp(\pm i2\pi y/\Lambda)$. To account for the non-ideality of the width modulation profiles and the rounded edges of the grating, the values of the grating $DC$ used in the simulations were slightly adjusted (in the range 70-82%) to match the experimental results. Finally, the values of the coupling coefficient were calculated as (1, 2):

$$\kappa(\lambda) = \frac{2\pi}{\lambda} \frac{\iint_{-\infty}^{+\infty} |F_0(x, z; \lambda)|^2 \, \delta n_1(x, z) \, dx \, dz}{\iint_{-\infty}^{+\infty} |F_0(x, z; \lambda)|^2 \, dx \, dz}$$

In the coupled-mode theory framework, the interaction between forward- and backward- propagating waves in a Bragg grating is described through slowly-varying envelopes, $A(y)$ and $B(y)$, respectively, which multiply the eigenmodes of the unperturbed waveguide. The compound response of the Bragg grating and phase-shifting sections can then be simulated by a transfer matrix method, which relates the complex envelopes of the forward and backward amplitudes at the input ($y = 0$) and output ($y = L$) of the overall structure through the relationship:

$$\begin{bmatrix} A_L \\ B_L \end{bmatrix} = M_{PSBG} \begin{bmatrix} A_0 \\ B_0 \end{bmatrix}$$

$M_{PSBG} = M_{BG}M_{PS}M_{BG}$ is the product of the 2x2 transfer matrices of the individual Bragg grating sections of length $\ell_{Bragg}$ ($M_{BG}$) and the phase shifting ($M_{PS}$) section of length $\delta L$ (Fig. 1b), which take the form:

$$M_m = \begin{bmatrix} a & b \\ c & d \end{bmatrix}, \quad m = BG, PS$$

with $s = \sqrt{\kappa^2 - (\Delta\beta/2)^2}$, $a = \cosh(s\ell_{Bragg}) - i\Delta\beta \sinh(s\ell_{Bragg})/2s$, $b = -i\kappa \sinh(s\ell_{Bragg})/s=-c$, $d = \cosh(s\ell_{Bragg}) + i\Delta\beta \sinh(s\ell_{Bragg})/2s$ for m = BG, while $a = e^{-i\varphi} = d^*$ and $b = c = 0$, for m = PS. In the case of a π-PSBG, the expression further simplifies into $\varphi_0 = \pi\lambda_0/(2\lambda)$.

The model allows to account for phase shifts different from π (i.e. $\delta L \neq \Lambda$), which move the transmission peak away from $\lambda_0$, as well as high-order Fabry-Perot-like multiwavelength structures induced by longer defects ($\delta L \gg \Lambda$), resulting in the appearance of multiple peaks within the transmission bandgap (3).

Cavity model.

As described in the text (equation (1)), attaining the narrowest linewidths requires maximizing both $Q_\alpha$ and $Q_\kappa$. This clearly pinpoints loss reduction as a key prerequisite for achieving ultra-narrow bandwidths. However, for the applications considered in this work, low losses have to be traded off against the higher grating strengths needed to minimize contributions from the term $Q_\kappa^{-1}$ to the bandwidth, while still keeping device footprints low. As outlined in the main text, a reasonable balance in the telecom band is obtained when $h$ (etching depth) is close to 300nm (for a wafer thickness of $H$ = 500nm) and the waveguide width $w_0 \sim$ 700 nm. This configuration yields selective $TE_{00}$-mode excitation (through properly dimensioned grating couplers), moderate propagation losses (α = 1.5 dB/cm) and relatively high grating strengths, realized at well-controlled modulation depths ($\delta w \sim 100$ nm) in very compact device formats (L~700μm).

Relying on the cavity model for long cavity devices ($R = \tanh^2(\kappa \ell_{Bragg})$ and equation (2) in the main text) one can estimate the bandwidth (and thus the Quality factor) of long cavity devices. The following plot (Figure S1) was obtained assuming propagation losses of 2dB/cm and relying on the data on the coupling coefficient κ extrapolated from the experimental measurements on π-PSBG previously fabricated. The typical coupling coefficient amounts to ~$10^2$ cm$^{-1}$. In the figure N represents the number of periods (i.e., ultimately, the device length). The plot shows the estimated quality factors for long cavity devices. We can see how:

a. a longer cavity leads to a higher Quality factor thanks to the mitigation of non-perfect coupling from the Bragg distributed mirrors.

b. The longer the device (i.e. the higher N) the higher the coupling itself, thanks to stronger mirrors (indeed the total coupling depends on the product $(\kappa \ell_{Bragg})$).

It can be seen (green circle) that there is a whole working region where the threshold of $10^5$ can be exceeded. This shows how long cavity devices are intrinsically more resilient against fabrication imperfections and more reproducible with respect to π-PSBG where a quality factor of $10^5$ can be reached (for example when $\alpha = 1.5 dB/cm$), only at critical coupling as explained in the main text. The high sensitivity to imperfections in the fabrication process may be detrimental for the performance of π-PSBG. On the other hand, the fact that π-PSBG are so sensitive to changes in the optical properties of the waveguide ensures that they perform excellently as sensors.

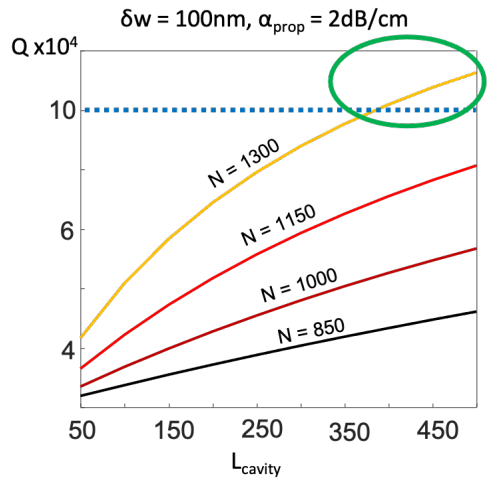

Figure S1. Simulations of the Quality factor for long cavity devices (L$_{cavity}$ in μm). N represents the number of periods. Higher N (i.e. longer devices) lead to a higher coupling to the cavity and thus to a higher Q. Fixed N, a longer cavity makes the peaks less sensitive to non-perfect coupling and allows to achieve a higher Q. Exceeding a cavity length of 450 μm would lead to a FSR smaller than 1nm and for this, that working region was not explored.

2. ELECTROSTATIC TUNING

The electrostatic properties of the waveguides were also investigated looking for the best compromise in terms of tuning efficiency vs. opt. loss contributions from the electrodes and slab thickness. The field distribution for the electrostatic field and for the optical mode were calculated using a commercial software (COMSOL, Wave Optics and COMSOL, Electrostatic). See also Figure S2. Then a simple model accounting for the linear electrooptic effect in waveguides was used.

Since $\lambda_0 = 2\bar{N}_0 \Lambda$ applying an external voltage (i.e. changing the effective refractive index), we have that the change in the resonance frequency reads:
$$d\lambda_0(V) = 2\Lambda\, d\bar{N}(V)$$
where
$$d\bar{N} = \frac{1}{2} n_o^3\, r_{33}\, \Gamma\, \frac{V}{g}$$

Here $n_o$ is the ordinary refractive index of LN, $r_{33}$ is the electrooptic coefficient to be accounted for given the orientation of the fabricated waveguides, $V$ is the applied voltage drop, $g$ represents the gap between the electrodes and $\Gamma$ is the overlap integral between the electrostatic field and the modulus square of the optical one (the $TE_{00}$). The simulation highlights how a gap smaller than 3μm would lead to optical losses, and that the presence of a slab below the ridge waveguide also improves the overlap (4).

In the implementation the thickness of the dielectric slab was chosen to be 200nm for a wafer thickness of 500nm. Results from the simulation and from the electrostatic measurements are shown in Figure S2.

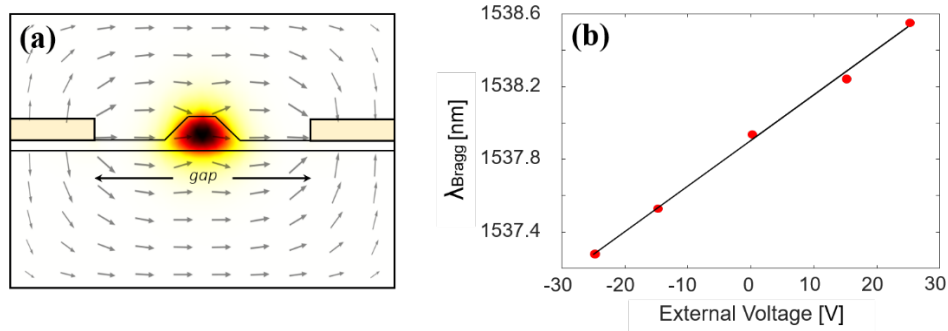

Figure S2. (a) Cross-sectional view of the waveguide with computed $TE_{00}$-mode electric field distribution at λ=1550 nm, for $w_0$=650nm, H=500nm, h=360nm and electrostatic field when the gap between the electrodes is 4 μm. Such simulations allowed the extraction of the overlap factor $\Gamma$. (b) Measured π-PSBG peak transmission wavelength as a function of the applied voltage. The linear fit (solid line) yields a spectral tunability $\sigma_\lambda = 25.1\ pm/V$.

3. π- PHASE SHIFTED BRAGG GRATINGS

Reaching experimentally the optimal working point for narrow bandwidth and critical coupling in a π-PSBG is by no means trivial, due to practical tradeoffs between waveguide loss ($\alpha$) and grating coupling strength ($\kappa$). For a given waveguide width $w_0$, $\kappa$ is controlled by the sidewall modulation amplitude: a deep modulation $\delta w$ leads to a stronger grating coupling, yet at the expense of scattering losses (5). Similarly, for a fixed value of $\delta w$, a larger average waveguide width $w_0$ lowers propagation losses but reduces the mode overlap with the grating, hence $\kappa$ (6). These (and other) tradeoffs are summarized in the following table:

|  | Effect on coupling | Effect on losses |
| --- | --- | --- |
| Larger modulation ($\delta w$) | Higher coupling strength since a higher $\delta w$ brings to a higher reflectivity of the Bragg mirror | Higher scattering losses due to the sidewall modulation |

| Larger width ($w_0$) | Lower coupling strength: the wider the waveguide, the lower the effect of dispersion | Lower losses due to lower mode confinement |
| --- | --- | --- |
| Longer device (L) | Higher coupling strength, since this is directly proportional to L | Higher losses related to the propagation losses and, in case of high modulation, also higher impact of scattering losses |
| Wafer thickness | Better coupling on a thinner wafer (e.g. 300 nm thin film LNOI), leading to a fully modulated waveguide with no slab | Lower losses on a thicker wafer (e.g. 500nm thin film LNOI) in the presence of a LN slab below the modulated nanowire reducing the vertical confinement |
| PMMA cladding rather than air cladding | Higher coupling because of the effect of "mode pulling" towards the top (and the sides) of the waveguide | Lower propagation losses thanks to lower index mismatch and lower confinement. Higher scattering losses due to the "pulling effect" that increases the coupling strength |

The impact of such tradeoffs on the response of the LNOI π-PSBG devices was systematically studied with theory and experiments. The next section summarizes the main findings.

4. IMPACT OF SIDEWALL MODULATION AMPLITUDE

Figure S3 illustrates the evolution of the π-PSBG spectral response with increasing amplitudes of the sidewall modulation $\delta w$. It shows experimental data (dots) from a waveguide set associated with the same unperturbed waveguide structure {$w_0, H, h$}, whose propagation losses in the absence of modulation amounted to $\alpha$ = 2.9 dB/cm (independently assessed by cut-back measurements). The BG period of these devices is fixed ($\Lambda$ =435nm).

Consistently with expected trends for the undercoupled cavity regime ($Q_\kappa < Q_\alpha$), the experimental spectra in Fig. S1a-d show a negligible peak power penalty ($\delta P \leq 0.9$dB) and a progressive narrowing of the transmission linewidth $\delta\lambda$, from 1.38nm to 113pm, as $\delta w$ is increased from ~80 nm to ~360 nm. Increasing values of the latter result also in a progressive red-shift of the resonance wavelength $\lambda_0$ and an increase of the filter extinction ratio $ER$. For $\delta w$ < 200 nm the experimental spectral responses are in full agreement with numerical simulations based on the model exposed in the first section of this Supplementary with $\alpha$ = 2.9 dB/cm: see plots in Fig. 3a-c as solid lines. When $\delta w$ is further increased (Fig. S3d-h) $ER$ and $\lambda_0$ keep increasing in agreement with theoretical predictions. However, the measured filter bandwidth $\delta\lambda$ (related to the loaded $Q$) does not appear to follow the expected trend and attain critical coupling. Such discrepancies are to be attributed to scattering losses due to the deep modulation.

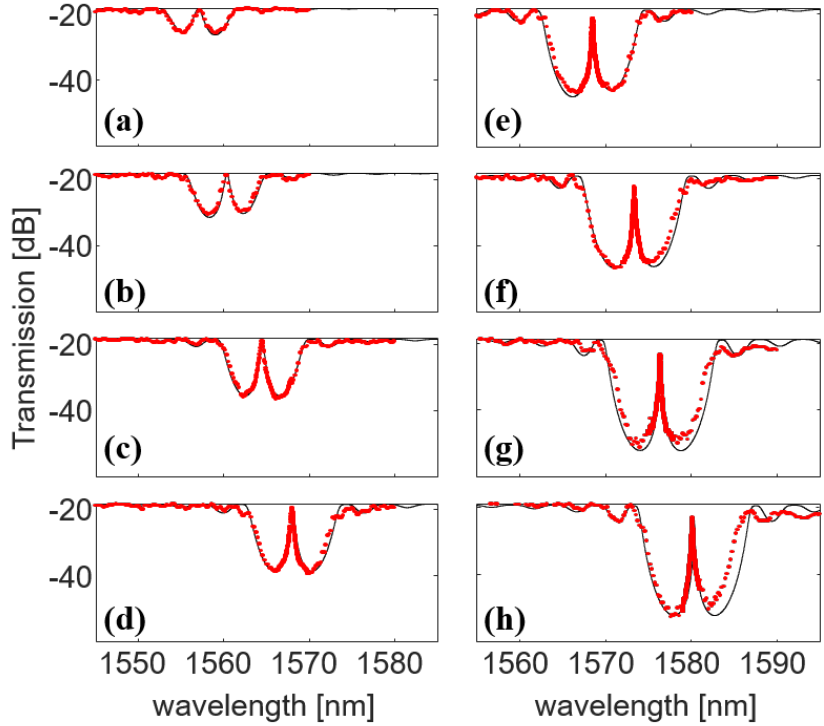

Figure S3. Measured (red dots) and calculated (solid lines) transmission spectra for a p-PSBG set with $w_0$ = 450 nm, $H$ = 500 nm, $h$ = 360 nm, $L$ = 435 nm and different grating modulation amplitudes $\delta w$: *a*) 82 nm, *b*) 125 nm, *c*) 178 nm, *d*) 230 nm, *e*) 243 nm, *f*) 280 nm, *g*) 320 nm, *h*) 364 nm, resulting in 3dB bandwidths $\delta\lambda$ = 1.38nm, 1.04nm, 380pm, 270pm, 212pm, 159pm, 113pm, 114pm, respectively. Duty cycles extrapolated through the fit: *a*) 71%, *b*) 74%, *c*) 78%, *d*) 80%, *e*) 78%, *f*) 80%, *g*) 80%, *h*) 82%.

As highlighted with reference to waveguide Bragg grating devices in SOI(3) the high sensitivity of the π-PSBG linewidth to losses allows a very accurate quantification of additional grating scattering terms in the fabricated waveguides. The extrapolated values of $\alpha_s$ for the experimental data set of Figure S3 are summarized in Figure S4.

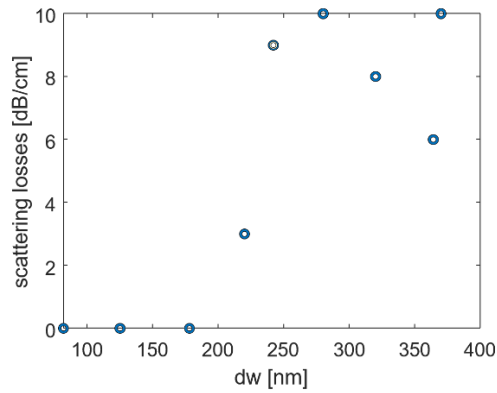

Figure S4. Plot of the measured scattering losses. For shallow modulation, scattering losses can be completely neglected. When the modulation becomes deeper, they become the dominant contribution to losses.

5. EFFECTIVE CAVITY LENGTH

Notwithstanding the fact that every device of this kind can be intuitively described as an on chip linear cavity between two Bragg mirrors, Equation (2) in the main text, used to evaluate the bandwidth for long cavity devices, cannot be used in the case of π-PSBG. As a matter of fact, the nature of Bragg mirrors leads to a non-negligible penetration of the radiation into the sidewall modulated portion of the waveguide from the cavity. For this reason, we define an effective cavity length: $L_{effective} = L_{cavity} + 2L_{penetration}$. The formula stems from a model (2) where losses are concentrated in cavity and mirrors are perfect. Therefore it is applicable only as long as the greatest contribution to intracavity losses comes from propagation losses, i.e. as long as $\alpha_0 L_{cavity} > 2(\alpha_0 + \alpha_s)L_{penetration}$. Since in our work we chose a modulation depth compatible with low scattering losses, this means that one needs $L_{cavity} > L_{penetration}$ to apply a Fabry Perot-like model.

Recalling the expression for the Free Spectral Range for a certain effective cavity length: $\Delta\lambda_{FSR} = \lambda_0^2 c/(2L_{effective} N_g)$, one can estimate the value of the penetration depth for the long-cavity devices presented in Figure 5 of the main text. Doing so we get: $L_{penetration} = 78.8 \mu m \pm 3.9 \mu m$, calculated based on the experimental results of all the long cavity devices fabricated on a same chip sharing the same geometrical parameter, up to the grating length with the ones in Fig.5. Such a penetration depth is larger but usually comparable with $L_{cavity}$ for our long cavity devices. As a consequence, Equation (2) proved to be a very useful tool to understand why the longer the cavity, the higher the Quality factor, but cannot be used for exact quantitative analyses.

Moreover, the calculated $L_{penetration}$ is much larger than the modulation period Λ. For this reason the Fabry Perot formula for the bandwidth cannot be applied at all to π-PSBG. Therefore an exact quantification of the grating response for π-PSBG can be obtained only relying on the numerical model described previously.

AUTHOR INFORMATION

Corresponding Author

* gallo@kth.se